\documentclass[a4paper, amsfonts, amssymb, amsmath, reprint, showkeys, nofootinbib, superscriptaddress]{revtex4-1}
\usepackage[english]{babel}
\usepackage[utf8]{inputenc}
\pdfoutput=1
\usepackage{amsthm}
\usepackage{amsmath}
\usepackage{mathtools}
\usepackage{physics}
\usepackage{xcolor}
\usepackage{graphicx}
\usepackage[left=18mm,right=18mm,top=35mm,columnsep=15pt]{geometry} 
\usepackage{adjustbox}
\usepackage{placeins}
\usepackage[T1]{fontenc}
\usepackage{balance}
\usepackage{csquotes}
\usepackage[pdftex, pdftitle={Article}, pdfauthor={Author}]{hyperref} 
\usepackage{floatrow, siunitx, booktabs}
\usepackage{color, colortbl}
\DeclareFloatFont{tiny}{\small}
\long\def\/*#1*/{}

\makeatletter 
\renewcommand{\thetable}{\arabic{table}}
\renewcommand{\fnum@figure}{\textbf{Fig.\ \thefigure}}
\renewcommand{\fnum@table}{\textbf{Table \thetable}}
\makeatother

\begin{document}

\title{Boosting the efficiency of Smith-Purcell radiators using nanophotonic inverse design}

\author{Urs Haeusler\textsuperscript{1,2,$\dagger$,*}}
\author{Michael Seidling\textsuperscript{1,$\dagger$,*}}
\author{Peyman Yousefi\textsuperscript{1,3}}
\author{Peter Hommelhoff\textsuperscript{1,*}}

\noaffiliation

\affiliation{Department Physik, Friedrich-Alexander-Universität Erlangen-Nürnberg (FAU), Staudtstraße 1, Erlangen 91058, Germany}
\affiliation{Present address: Cavendish Laboratory, University of Cambridge, JJ Thomson Avenue, Cambridge CB3 0HE, UK}
\affiliation{Present address: Fraunhofer-Institut f\"ur Keramische Technologien und Systeme IKTS, \"Aussere N\"urnberger Strasse 62, Forchheim 91301, Germany
\\ \ \\
\textsuperscript{$\dagger$}\,These authors contributed equally to this work.
\\
\textsuperscript{*}\,email: uph20@cam.ac.uk; michael.seidling@fau.de; peter.hommelhoff@fau.de
\\ \ \\
}


\begin{abstract}
The generation of radiation from free electrons passing a grating, known as Smith-Purcell radiation, finds various applications including non-destructive beam diagnostics and tunable light sources, ranging from terahertz towards X-rays. So far, the gratings used for this purpose have been designed manually, based on human intuition and simple geometric shapes. Here we apply the computer-based technique of nanophotonic inverse design to build a $1400\,\mathrm{nm}$ Smith-Purcell radiator for sub-relativistic $30\,\mathrm{keV}$ electrons. We demonstrate that the resulting silicon nanostructure radiates with a 3-times-higher efficiency and 2.2-times-higher overall power than previously used rectangular gratings. With better fabrication accuracy and for the same electron-structure distance, simulations suggest a superiority by a factor of 96 in peak efficiency. While increasing the efficiency is a key step needed for practical applications of free-electron radiators, inverse design also allows to shape the spectral and spatial emission in ways inaccessible with the human mind.

\end{abstract}

\maketitle


The Smith-Purcell effect describes the emission of electromagnetic radiation from a charged particle propagating freely near a periodic structure. The wavelength $\lambda$ of the far-field radiation follows \cite{smith1953visible}
\begin{equation}
\lambda = \frac{a}{m} \big ({\beta}^{-1} - \cos{\theta} \big ), \label{eq:Smith-Purcell relation}
\end{equation}{}
where $a$ is the periodicity of the structure, $\beta = v/c$ the velocity of the particle, $\theta$ the angle of emission with respect to the particle propagation direction, and $m$ the integer diffraction order.

The absence of a lower bound on the electron velocity in equation (\ref{eq:Smith-Purcell relation}) makes Smith-Purcell radiation (SPR) an interesting candidate for an integrated, tunable free-electron light source in the low-energy regime \cite{adamo2009light, liu2017integrated, massuda2018smith, yang2018maximal, roques2019towards, ye2019deep, shentcis2020tunable}. While the power efficiency of this process is still several orders of magnitude smaller than conventional light sources, it can be enhanced by superradiant emission from coherent electrons \cite{gover2019superradiant}. For this, pre-bunching of the electrons is a possible avenue \cite{ishi1995observation, korbly2005observation, zhang2015enhancement, pan2018spontaneous}, but also self-bunching due to the interaction with the excited nearfield of the grating is observed above a certain current threshold \cite{urata1998superradiant, andrews2004gain, kumar2006analysis}. The use of coherent electrons is particularly interesting in combination with resonant structures, such as near bound states in the continuum \cite{hsu2016bound, yang2018maximal}.

Even in the regime of incoherent electrons, Smith-Purcell radiation can be greatly enhanced by optimizing beam parameters (velocity and diameter) and grating properties (material and shape). The latter are generally limited by the chosen method of fabrication. Typical gratings for the generation of near-infrared, visible or ultraviolet light are fabricated by reactive-ion etching or focused-ion-beam milling of silicon or fused silica \cite{adamo2009light, liu2017integrated, yang2018maximal, massuda2018smith, roques2019towards, ye2019deep, karnieli2021smith}. Coating the grating with a metal such as gold, silver or aluminium can lead to plasmonic enhancement \cite{chuang1984enhancement, so2015amplification, lai2017generation, kaminer2017spectrally}.

While a simple rectangular grating has been the most common choice in Smith-Purcell experiments, other designs have been investigated, such as metasurfaces and aperiodic structures \cite{wang2016manipulating, remez2017spectral, kaminer2017spectrally, su2019manipulating, su2019complete, wang2020vortex, karnieli2021smith, yang2021observation}. Here, we explored the optimization technique of nanophotonic \textit{inverse design} \cite{piggott2015inverse, hughes2017method, molesky2018inverse, sapra2020chip, dahan2021imprinting} to generate SPR much more efficiently. In contrast to other photonic designs created manually and optimized for a small set of parameters, inverse design finds an optimal design without any prior knowledge of its shape, purely based on the desired performance.

\begin{figure*}
  \centering
  \includegraphics{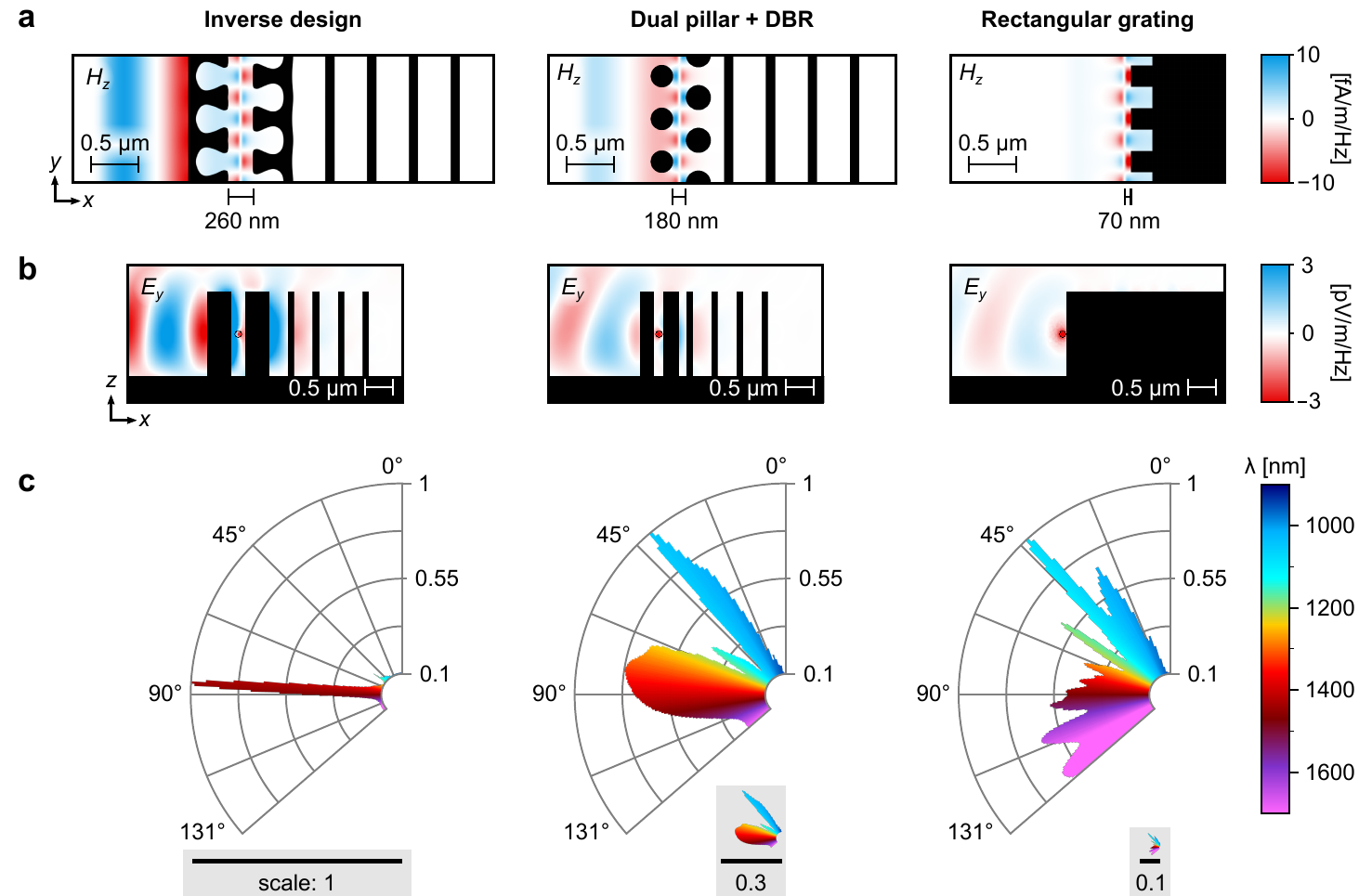}
  \caption{\textbf{Radiation characteristics.} The inverse-designed structure (left) is compared to a dual-pillar design with distributed Bragg reflector (DBR, center) and a rectangular grating (right). \textbf{a}, Magnetic field profile at $\lambda = 1400\,\mathrm{nm}$ obtained from 2D-FD simulation, as used in the optimization process. \textbf{b}, 3D-FD simulation of the electric field profile at $1370\,\mathrm{nm}$ as seen from the perspective of the electron beam (red dot). \textbf{c}, Emission spectrum calculated from 3D-FD simulation. The inverse-designed structure radiation is sharply peaked, both spectrally and in direction, whereas the other two structures emit much more broadbandly, spectrally and spatially.}
  \label{fig1}
\end{figure*}

We applied the technique to maximize SPR from $30\,\mathrm{keV}$ electrons ($\beta = 0.328$) passing through a silicon nanostructure and radiating around $\lambda = \SI{1.4}{\micro m}$ in the transverse direction ($\theta = 90^\circ$). The resulting nanostructure forms an asymmetric cavity around the electron beam, which leads to a highly concentrated emission into a well-defined direction. We compared the emission characteristics to those of a structure with a double row of pillars and a distributed Bragg reflector as well as that of a rectangular grating (Fig.\ \ref{fig1}). Like most previously used gratings, these radiate broadbandly, both spectrally and spatially (Fig.\ \ref{fig1}c). This impedes their application as a light source because part of the electron energy is converted to radiation that cannot be collected or is spectrally irrelevant. By contrast, the here presented inverse design can resolve these problems with unprecedented efficiency.

\section*{Design}

The inverse design optimization was carried out via an open-source Python package \cite{hughes2019forward} based on a 2D frequency-domain (FD) simulation. At the center of the optimization process is the objective function $G$, which formulates the desired performance of the design, defined by the design variable $\phi$ (Methods). Here, we aimed for maximum radiation in negative $x$-direction at the design angular frequency $\omega$ corresponding to $\lambda = \SI{1.4}{\micro m}$ (Fig.\ \ref{fig1}). To this end, the Poynting vector $S$ was numerically measured in the far field of the structure and integrated over one period $a$, giving the objective function
\begin{equation}
    G(\phi) = - \int_0^a dy \, S_x(x_\mathrm{far \, field}, y). \label{eq:ObjFctSPR}
\end{equation}

The resulting design is depicted in Fig.\ \ref{fig1}a and reveals two gratings on each side of the vacuum channel, which are similar in shape but $\pi$-phase shifted with respect to each other. The back of the double-sided grating results in a structure which resembles a distributed Bragg reflector (DBR). This way, the radiation to the left is 469-times higher than to the right.

\begin{figure*}
  \centering
  \includegraphics{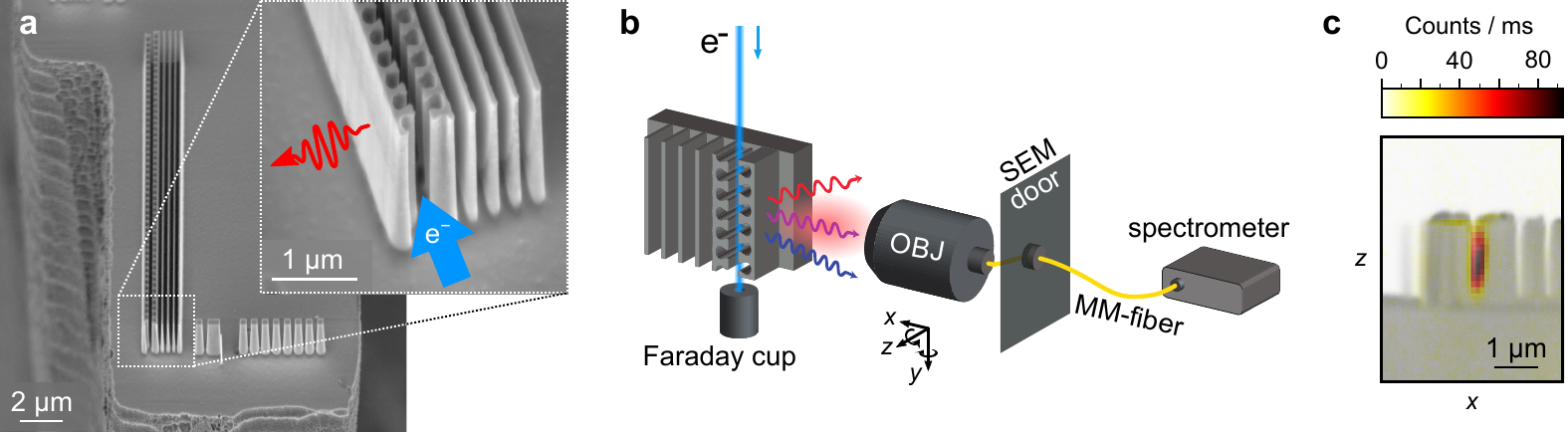}
  \caption{\textbf{Fabricated inverse-designed structure and experimental setup.} \textbf{a}, SEM image of the inverse-designed silicon nanostructure. It sits on a 50-\SI{}{\micro m}-high mesa to provide clearance for the electron beam and photon collection. \textbf{b}, Setup used to measure Smith-Purcell radiation. The electron beam from an SEM is focused into the structure, and the current is measured with a Faraday cup afterwards. An objective (OBJ) on a 5D alignment stage collects the generated photons. A multimode (MM) fiber guides the light outside the SEM, where it is spectrally resolved with a spectrometer. \textbf{c}, Photocounts as a function of the electron beam position. It is overlaid on an SEM image, revealing strongest radiation at the center of the electron channel.}
  \label{fig2}
\end{figure*}

A 200-period-long version of the inverse-designed structure was fabricated by electron beam lithography ($100\,\mathrm{kV}$) and cryogenic reactive-ion etching of $1$-$5\,\mathrm{\Omega cm}$ phosphorus-doped silicon to a depth of \SI{1.3(1)}{\micro m} \cite{yousefi2019dielectric}. The surrounding substrate was etched away to form a 50-\SI{}{\micro m}-high mesa (Fig.\ \ref{fig2}a). We note that unlike in most previous works the etching direction is here perpendicular to the radiation emission, enabling the realisation of complex 2D geometries.

The radiation generation experiment was performed inside a scanning electron microscope (SEM) with an $11\,\mathrm{nA}$ beam of $30\,\mathrm{keV}$ electrons. The generated photons were collected with an objective (NA 0.58), guided out of the vacuum chamber via a 300-\SI{}{\micro m}-core multimode fiber, and detected with a spectrometer (Fig.\ \ref{fig2}b and Methods).

\section*{Results}
We compare the emission characteristics of the inverse-designed structure to two other designs: Firstly, a rectangular 1D grating with groove width and depth of half the periodicity $a$, similar to the one used in \cite{roques2019towards, szczepkowicz2020frequency}. And secondly, a \textit{dual pillar} structure with two rows of pillars, $\pi$-phase shifted with respect to each other, and with a DBR on the back. This design was successfully used in dielectric laser acceleration, the inverse effect of SPR \cite{peralta2013demonstration, breuer2013laser, leedle2015dielectric, hughes2017method, yousefi2019dielectric, sapra2020chip, dahan2021imprinting}. It further represents the man-made design closest to our result of a computer-based optimization.

Figure \ref{fig2}c shows the photon count rate as a function of the electron beam position. Maximum photon count rate is observed when focusing the beam into the channel of the inverse design structure at medium height. The spatial confinement in vertical direction points at the presence of a confined mode, as found in cavities. By contrast, Fig.\ S1 reveals a non-resonating nature of the other two structures, with only slight dependence on the beam height.

\begin{figure*}
    \centering
    \includegraphics{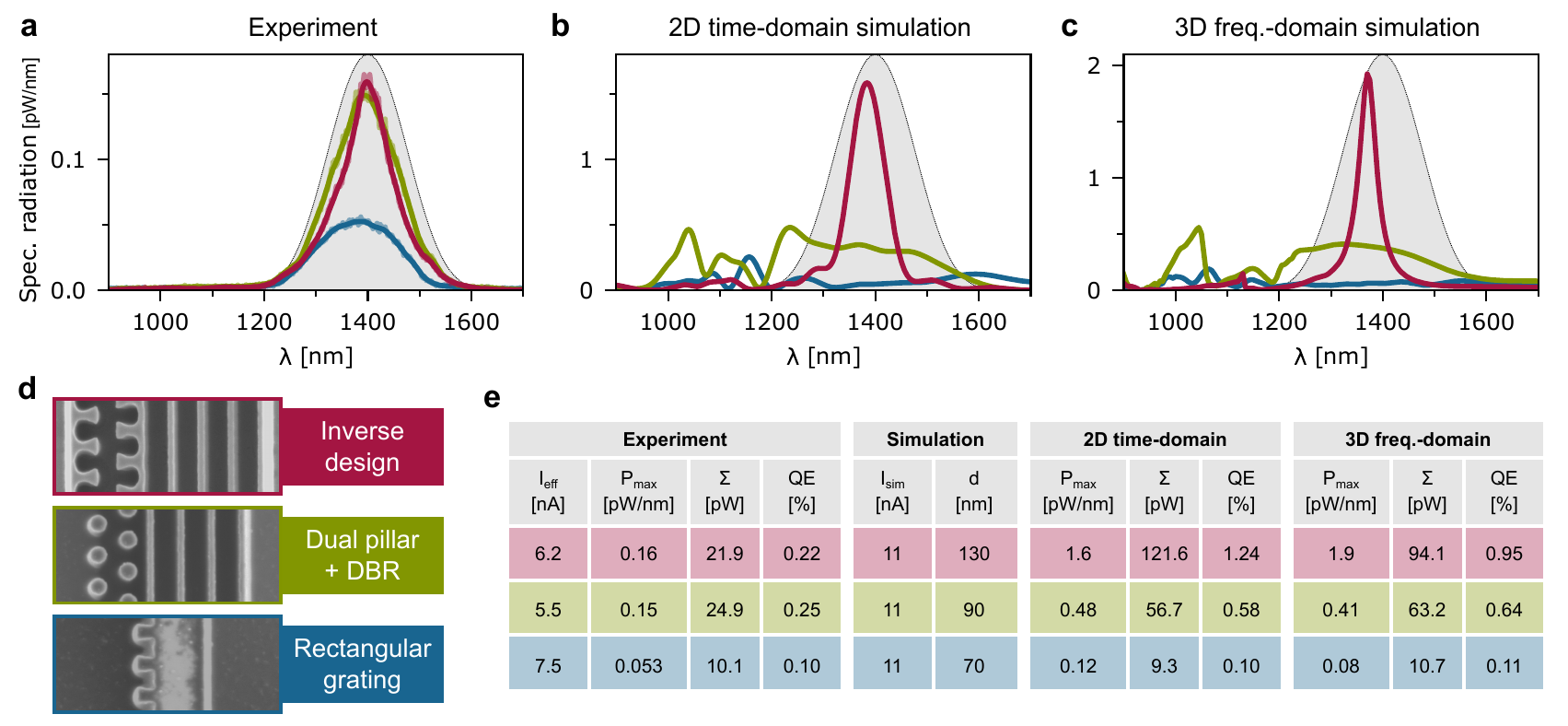}
    \caption{\textbf{Comparison of radiation spectra}. Inverse design (red), dual pillar structure (green) and rectangular grating (blue) are shown in front of the Gaussian filter (grey) that accounts for the limited collection range. \textbf{a}, Measurement. \textbf{b}, 2D time-domain simulation (2D-TD). \textbf{c}, 3D frequency-domain simulation (3D-FD). \textbf{d}, SEM pictures of the three structures. \textbf{e}, Table showing the peak spectral radiation density $P_\mathrm{max}$ in the experimentally accessible range from $1200\,\mathrm{nm}$ to $1600\,\mathrm{nm}$, the total power $\Sigma$ and quantum efficiency $QE$, both calculated using the Gaussian filter. The effective current $I_\mathrm{eff}$ accounts for electron loss within the structure (Fig.\ S2). While the inverse design performed similarly to the dual pillars in the experiment, the simulations suggest higher peak and overall efficiency. This superiority is much stronger when compared to the rectangular grating.}
    \label{fig3}
\end{figure*}

The efficiency of a design is quantified by comparing three different figures of merit: the peak spectral radiation density (pW/nm), the total radiation (pW), and the quantum efficiency ($\%$), defined as the number of photons generated per electron. All three quantities are determined in the experimentally accessible window, which is limited by the numerical aperture of the fiber. Its angular acceptance window acts as an effective spectral filter with a Gaussian shape centered around $1400\,\mathrm{nm}$ and a full width at half maximum of $175\,\mathrm{nm}$ (Fig.\ \ref{fig3} and Methods).

The measurements in Fig.\ \ref{fig3}a show overall a similar performance of the inverse-designed structure and the dual-pillar structure. In terms of overall power, the inverse design is with $21.9\,\mathrm{pW}$ around 12\% weaker than the dual pillars. This can be understood by the larger channel width of $260\,\mathrm{nm}$ compared to the $180\,\mathrm{nm}$ of the dual pillars (Fig.\ \ref{fig1}a). By contrast, the rectangular grating is single sided, and the beam was steered as closely as possible to the grating to yield maximum radiation. Even then, the inverse design radiated 2.2(1)-times as strong as the rectangular grating. The superiority becomes even more pronounced when looking at the peak spectral radiation density. The inverse design reaches $0.16\,\mathrm{pW/nm}$ at $1385\,\mathrm{nm}$, which is 3.0(1)-times as high as that of the rectangular grating. It also surpasses marginally the dual pillar peak efficiency. This is a first indicator for the narrowband emission of the inverse design, in contrast to the broadband emission of the other two designs (Fig.\ \ref{fig1}c).

For further study of the different structures, we performed 2D time-domain and 3D frequency-domain simulations. While both time and frequency domain are in principal legitimate ways to calculate the radiation spectrum from single electrons, they differ in computational complexity and precession. The time-domain simulation (Fig.\ \ref{fig3}b and Fig.\ S4) can capture the instantaneous response to a structure of finite length. This is computationally expensive because the field of the entire grating needs to be calculated at each point in time. The frequency-domain simulation (Fig.\ \ref{fig3}c) on the other hand calculates the radiation density at each frequency of the spectrum. This is computationally less complex because it is sufficient to considers a single unit cell with periodic boundaries, which allowed us to perform 3D simulations. It can therefore take into account the limited height of the electron beam and the structure, which is on the order of the wavelength. This is particularly relevant here because the inverse design yielded a double-sided grating that forms a resonator. The mirrors of the resonator are plane parallel and therefore do not form a stable resonator.

Both 2D time-domain and 3D frequency-domain simulations show similar results. For the inverse design, they predict a total radiation of $108(14)\,\mathrm{pW}$, a quantum efficiency of $1.1(2)\%$ and a peak spectral radiation density of $1.8(2)\,\mathrm{pW/nm}$. In terms of total power, this corresponds to an increase by 80\% compared to the dual pillar design and a colossal boost of 980\% with respect to the rectangular grating. The contrast in terms of peak efficiency within the experimentally accessible range from $1200\,\mathrm{nm}$ to $1600\,\mathrm{nm}$ is even more drastic. It reaches an increase by 290\% compared to the dual pillars and 1650\% relative to the rectangular grating.

\begin{table*}
  \centering
  \includegraphics{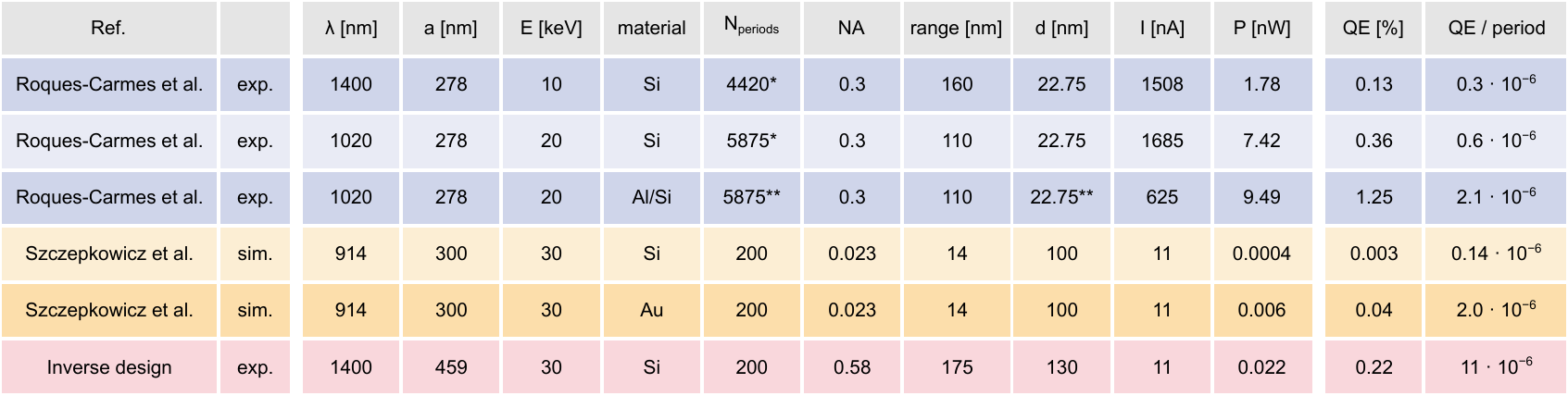}
  \caption{\textbf{Comparison to literature.} The inverse design structure is compared to rectangular gratings used in a similar experiment by Roques-Carmes \textit{et al.} \cite{roques2019towards} and simulated by Szczepkowicz \textit{et al.} \cite{szczepkowicz2020frequency}. Their gratings were made of silicon (Si), aluminium coated silicon (Al/Si), and gold (Au). In case of \cite{roques2019towards}, the number of periods was stated as (*) the effective number of periods based on the simulated interaction length or (**) not explicitly stated. The collection ranges in the experiments were estimated as full widths at half maximum of the measured spectra. While the many different parameters jeopardize a direct comparison, it is clear that the inverse design structure at this work is far superior when it comes to the probably most relevant quantity: the (quantum) efficiency per period. The inverse designed structure is at least a factor of 5 better.}
  \label{tab1}
\end{table*}

\section*{Discussion}
Comparing the measured emission spectrum of the inverse design to its simulated profile shows that the observed emission was not as powerful and spectrally broader. We identify two causes: Firstly, the electron beam current deteriorates as the beam diverges, where electrons hit the boundaries of the channel and are lost. By measuring the current after the structure, we determined an effective current $I_\mathrm{eff}$ for each design (Fig.\ \ref{fig3}e and Fig.\ S2). The effective current is smallest for the dual pillar design, which has the narrowest channel, and largest for the single-sided rectangular grating.

\begin{figure}
    \centering
    \includegraphics{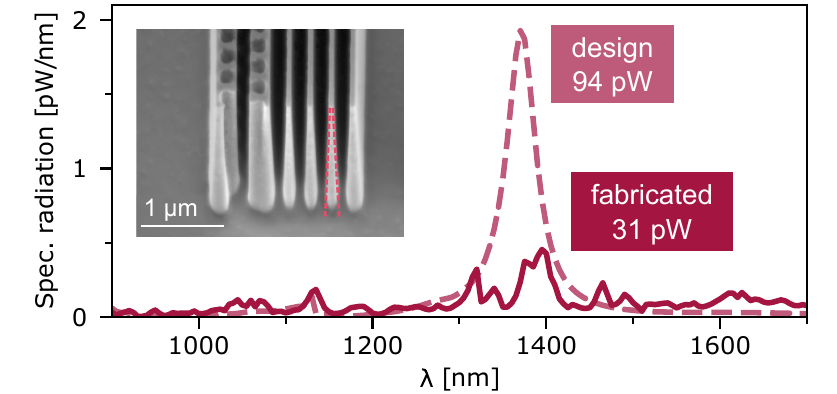}
    \caption{\textbf{Effect of non-vertical etch.} The SEM image (inset) shows the non-ideal etching result with slightly conical instead of rectangular shapes - presumably caused by an imbalance of the $O_2$-$SF_6$ gas mixture during etching. The conical shape reduces the quality factor of the inverse-designed structure and leads to a degradation of efficiency and spectral broadening. This can be seen from the simulated (3D-FD) spectral radiation density of the designed and fabricated structures as well as their integrated powers over the Gaussian-shaped collection range.}
    \label{fig4}
\end{figure}

Another factor that reduces the efficiency of the inverse-designed structure are the deviations of the fabricated structure from its design. Figure \ref{fig4} shows that the structure was not perfectly vertically etched but has slightly conical features. This leads to a reduction of the quality factor of the inverse-designed structure, which is reflected in a less powerful ($-67\%$) and more broadband emission of radiation. By contrast, the efficiencies of the dual pillar structure and the rectangular grating are expected to be less affected by conical features due to their lack of pronounced resonance.

Finally, we note that the inverse design was operated with the beam at the center of the 260-$\mathrm{nm}$-wide channel, whereas the rectangular grating worked at minimal beam-structure distance for maximal efficiency. The simulations assumed the distance $d = 70\,\mathrm{nm}$, whereas Fig.\ S3 suggests that the actual distance in the experiment was $58\,\mathrm{nm}$. Would the rectangular grating have been operated at $d=130\,\mathrm{nm}$, the simulations predict that the inverse design could improve peak spectral radiation density and overall radiation power by factors of 96 and 42, respectively.\\

It is interesting to relate the quantum efficiency of our inverse-designed structure to that of other silicon gratings reported elsewhere (Table \ref{tab1}). Roques-Carmes et al.\ \cite{roques2019towards} state a quantum efficiency of $0.13\%$ in a similar experiment at $\lambda = 1400\,\mathrm{nm}$. Although their interaction length was 13-times longer, and the distance to the grating was with $d=23\,\mathrm{nm}$ just a fifth of ours, the inverse-designed structure surpasses it with its quantum efficiency of $0.22\%$.

A similar conclusion can be drawn in comparison to the simulations of a silicon grating at $\lambda = 914\,\mathrm{nm}$ by Szczepkowicz et al.\ \cite{szczepkowicz2020frequency}. While the collection range in our experiment was 13-times broader, the here observed quantum efficiency was 78-times higher.

Comparing the experimental aspects, our approach of a double-sided grating poses higher demands on the beam quality than a single-sided grating. As such, the beam can only pass the grating if aligned parallel, and as shown, divergence limits the effective current for longer gratings. Moreover, our chosen method of fabrication restricts the grating region to the height of the structure and is predestined for in-plane collection. However, the collimated emission pattern achieved by our grating can simplify collection through fibers with small numerical aperture.

\section*{Conclusion}

In summary, our work introduces the technique of inverse design to the field of Smith-Purcell radiators. We presented a silicon-photonic nanostructure radiating in the NIR regime for sub-relativistic $30\,\mathrm{keV}$ electrons with a peak spectral density 3-times higher and overall power 2.2-times higher than a rectangular grating. With smaller fabrication inaccuracies and for the same beam-grating distance, simulations predict a 96-fold improvement in peak efficiency and 42-times-higher total power.

The superiority lies firstly in our approach of using a double-sided grating in combination with a DBR - an idea which was brought up in dielectric laser accelerators \cite{leedle2015dielectric, hughes2017method, yousefi2019dielectric} and became possible through recent advances in nanofabrication. Secondly, the inverse design algorithm suggests to use a closed scheme where the radiation is reflected at silicon boundaries. This leads to resonant enhancement and superradiance if used together with coherent electrons \cite{korbly2005observation, gover2019superradiant}.

Further improvement of the efficiency can be achieved by coating the dielectric grating with a metal, which has been reported to gain enhancement by a factor of 3 to 14 for rectangular gratings \cite{roques2019towards, szczepkowicz2020frequency}.

The advantage of inverse-designed Smith-Purcell radiators goes beyond just higher efficiency. We presented a structure that radiates in a spatially and spectrally well-defined direction. More generally, the versatility of the optimization technique allows to design the spectrum ($\omega$), spatial distribution ($\mathbf{r}$) and polarization ($\mathbf{e}$) of radiation by favoring one kind $\abs{\mathbf{e}\cdot\mathbf{E}(\mathbf{r}, \omega)}$ and penalizing others $-\abs{\mathbf{e}'\cdot\mathbf{E}(\mathbf{r}', \omega')}$ with possibly orthogonal polarization $\mathbf{e}'$. Lifting the periodicity constraint opens the space to complex metasurfaces, which would for example enable designs for focusing or holograms \cite{zheng2015metasurface, wang2016manipulating, remez2017spectral, kaminer2017spectrally, su2019manipulating, su2019complete, wang2020vortex, karnieli2021smith}.

Future efforts could also target the electron dynamics to achieve (self-)bunching and hence coherent enhancement of radiation. In that case, the objective function would aim at the field inside the electron channel rather than the far-field emission. This would favor higher quality factors at the cost of lower out-coupling efficiencies. However, direct inclusion of the electron dynamics through an external multi-physics package proves challenging as our inverse design implementation requires differentiability of the objective function with respect to the design parameters. Instead, one may choose to use an analytical expression for the desired electron trajectory or an approximate form for the desired field pattern.

\bibliography{main}

\section*{Methods}
\begin{figure}[hb]
  \centering
  \includegraphics{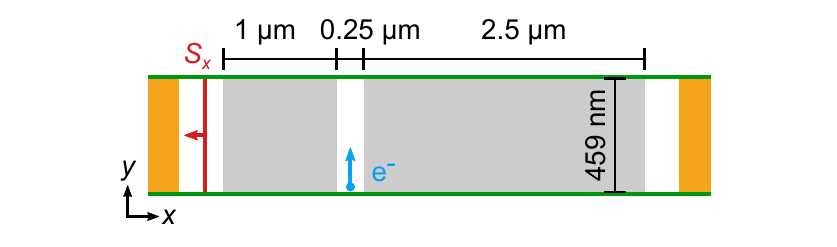}
  \caption{\textbf{Simulation cell for inverse design}. The structure was optimized over a region which extended \SI{1}{\micro m} to the left and \SI{2.5}{\micro m} to the right of a 250-nm-wide vacuum channel, ensuring free propagation of the electrons (blue arrow). Periodic boundaries (green) were applied in longitudinal direction, setting the periodicity to $a = 459\,\mathrm{nm}$, and perfectly matched layers (orange) were defined at each side of the simulation cell in transverse direction. The Poynting vector (red) in negative x-direction $S_x$ was calculated outside the design region and served as the objective function.} \label{figED1}
\end{figure}

\textbf{Inverse design.}
The inverse design optimization was carried out via an open-source Python package \cite{hughes2019forward} based on a 2D finite-difference frequency-domain (FDFD) simulation at the design angular frequency $\omega$ corresponding to $\lambda = \SI{1.4}{\micro m}$. The simulation cell used for this purpose is presented in Fig.\ \ref{figED1}.

The design $\varepsilon_\text{r}(\phi)$ was parametrized with the variable $\phi(\mathbf{r})$. Sharp features ($<100\,\mathrm{nm}$) in the design were avoided by convolving  $\phi(\mathbf{r})$ with a 2D circular kernel of uniform weight. Afterwards the convolved design $\tilde{\phi}$ was projected onto a sigmoid function of the form $\tanh{(\gamma \tilde{\phi})}$. This results in a close-to-binary design where the relative permittivity $\varepsilon_r(\mathbf{r})$ only takes the values of silicon ($\varepsilon_\text{r} = 12.2$) \cite{schinke2015uncertainty} or vacuum ($\varepsilon_\text{r} = 1$). We observed good results by starting the optimization with small values $\gamma = 20$ and slowly increasing $\gamma$ to 1000.

To improve the convergence of the algorithm, we enforced mirror symmetry in x-direction onto the design. This reflects the symmetry of SPR under $\theta = 90^\circ$ and reduces the parameter space by a factor of 2. Furthermore, we observed improved convergence when starting with a large grid spacing ($10\,\mathrm{nm}$), which is then slowly reduced to $3\,\mathrm{nm}$ as the optimization progresses.\\

\textbf{2D simulations.}
The source term of our 2D simulations is given by the current density of a line charge with density $q = -e/\Delta z$ traversing the structure with velocity $v$ along $\mathbf{\hat{y}}$.
The choice of the length $\Delta z$ is crucial to obtain meaningful intensities from a 2D simulation \cite{szczepkowicz2020frequency}. By choosing $\Delta z = \SI{1}{\micro m}$ throughout, we obtained 2D results that were on average only 14\% off the 3D values.
In transverse direction, we assumed a Gaussian charge distribution of width $\sigma_x=20\,\mathrm{nm}$ such that the spectral current density reads
\begin{align*}
    \mathbf{J}(\mathbf{r}, \omega) = \frac{q}{2 \pi} \cdot (2 \pi \sigma_x^2)^{-1/2} \cdot \mathrm{e}^{-x^2/2\sigma_x^2} \cdot \mathrm{e}^{-i k_\text{y} y} \, \mathbf{\hat{y}}
\end{align*}
with $k_y = \omega / v$. Using this expression, the electromagnetic field was calculated via Maxwell's equations for linear, non-magnetic materials. As this is a 2D problem, the transverse-electric mode $E_z$ decouples from the transverse-magnetic mode $H_z$, where only the latter is relevant here. A typical 2D-FDFD simulation took $1\,\mathrm{s}$ on a common laptop, and the algorithm needed about 500 iterations to converge to a stable maximum.\\

\textbf{Simulated radiation power.}
From the simulated electromagnetic field, we calculate the total energy $W$ radiated by a single electron per period $a$ of the grating. In the time domain, this would correspond to integrating the energy flux $\mathbf{S} (\mathbf{r}, t)$ through the area surrounding the grating over the time it takes for the particle to pass over one period of the grating. In the frequency domain, one needs to integrate $\mathbf{S} (\mathbf{r}, \omega)$ through the area around one period over all positive frequencies, that is \cite{szczepkowicz2020frequency},
\begin{align}
    W &= \int_S d\mathbf{A} \cdot \int_{0}^{\infty} d\omega \,\, \mathbf{S} (\mathbf{r}, \omega),\\
    \mathbf{S} (\mathbf{r}, \omega) &= 4 \cdot 2\pi  \,\, \Re{\frac{1}{2}\,\mathbf{E}(\mathbf{r}, \omega) \times \mathbf{H^*}(\mathbf{r}, \omega)},\nonumber\label{eq:spectralPoyntingVector}
\end{align}{}
where we chose a surface $\int_S d\mathbf{A} = - a \mathbf{\hat{x}} \int dz$ parallel to the grating as we were only interested in radiation in negative x-direction. For 2D simulations, the area $A=a \cdot \Delta z$ is determined by the assumed length $\Delta z$ of the line charge density $q=-e/\Delta z$ corresponding to one electron.\\

\textbf{Numerical instabilities.}
We observed that the optimization for a single frequency is very sensitive to numerical instabilities, which is why we optimized our design for multiple frequencies $\omega_i$ ($i=1,...,N$) simultaneously. A suitable objective function could be the sum over all $G(\phi, \omega_i)$, but we found that the $\min$-function
\begin{equation*}
    f_\text{obj}(\phi) = \min_i G(\phi, \omega_i)
\end{equation*}
was even more robust against numerical instabilities. Our design was optimized for the three $\omega$'s corresponding to $\lambda_{1,2,3} = 1350, 1400, 1450\,\mathrm{nm}$.\\

\textbf{Dual pillar design.}
The dual pillar design is inspired from \cite{yousefi2019dielectric}. Pillar radii and DBR thicknesses were optimized using the same gradient-based algorithm as for inverse design \cite{hughes2019forward}. Pillars that are $\pi$-phase-shifted with respect to each other are preferred over symmetric rows of pillars because they yield a stronger phase difference in $E_y$ and therefore stronger coupling to the far field.\\

\textbf{3D simulations.}
3D finite-element-method (FEM) frequency-domain simulations were performed in COMSOL to analyse effects originating from the finite height of structure and beam. The structures were assumed to be \SI{1.5}{\micro m} high on a flat silicon substrate (Fig.\ \ref{fig1}b). The spectral current density had a Gaussian beam profile of width $\sigma = 20\,\mathrm{nm}$:
\begin{align*}
    \mathbf{J}(\mathbf{r}, \omega) = \frac{-e}{2\pi} \cdot \left(2\pi \sigma^2 \right)^{-1} \mathrm{e}^{-(x^2+z^2)/2\sigma^2} \cdot \mathrm{e}^{-i k_\text{y} y} \, \mathbf{\hat{y}}.
\end{align*}

\textbf{Experimental setup.}
The experiment was performed within an \textit{FEI/Philips XL30} SEM providing an $11\,\mathrm{nA}$ electron beam with $30\,\mathrm{keV}$ mean electron energy. The structure was mounted to an electron optical bench with full translational and rotational control. The generated photons were collected with a microfocus objective Schäfter+Kirchhoff 5M-A4.0-00-S-Ti with a numerical aperture of $0.58$ and a working distance of $1.6\,\mathrm{mm}$. The objective can be moved relative to the structure with five piezoelectric motors for the three translation axes and the two rotation axes transverse to the collection direction. The front lens of the objective was shielded with a fine metal grid to avoid charging with secondary electrons in the SEM, which would otherwise deflect the electron beam, reducing its quality. The collected photons were focused with a collimator into a 300-\SI{}{\micro m}-core multimode fiber guiding the photons outside the SEM, where they were detected with a NIRQuest+ spectrometer.\\

\textbf{Collection range.}
The measured Gaussian spectrum from Fig.\ \ref{fig3}a can be explained by the limited numerical aperture of the collection fiber. Smith-Purcell radiation that is emitted in non-perpendicular direction is offset from the optical axis for collection. This leads to a loss in collection efficiency, which we modelled with the function $\exp{-2r^2/ (f \cdot \text{NA})^2}$, where $r$ is the offset measured at the collimator, $f = 12\,\mathrm{mm}$ is the focal length of the collimator and NA the numerical aperture of the fiber. We found good agreement with the experimental data for $\text{NA}=0.11$, which is below the 0.22 stated by the manufacturer and might have been a result of misalignment.

\section*{Acknowledgments}
We thank Ian A. D. Williamson for helpful support with the inverse design software. We thank R.\ Joel England and Andrzej Szczepkowicz for fruitful discussions about the design and simulations. We acknowledge continued technical support by the MPL cleanroom staff. \textbf{Funding:} This project has received funding from the Gordon and Betty Moore Foundation grants 4744 (ACHIP) and 5733 (QEMII) as well as ERC Advanced Grant 884217 (AccelOnChip).

\section*{Author contributions}
U.H., M.S., and P.H. conceived the project and prepared the manuscript. U.H. designed, and P.Y. fabricated the structures. M.S. and U.H. acquired and analysed data and performed simulations.

\section*{Competing interests}
The authors declare no competing interests.

\makeatletter 
\renewcommand{\fnum@figure}{\textbf{Figure S\thefigure}}
\renewcommand{\fnum@table}{\textbf{Table S\thetable}}
\setcounter{figure}{0}
\setcounter{table}{0}
\makeatother

\begin{figure*}
  \centering
  \includegraphics{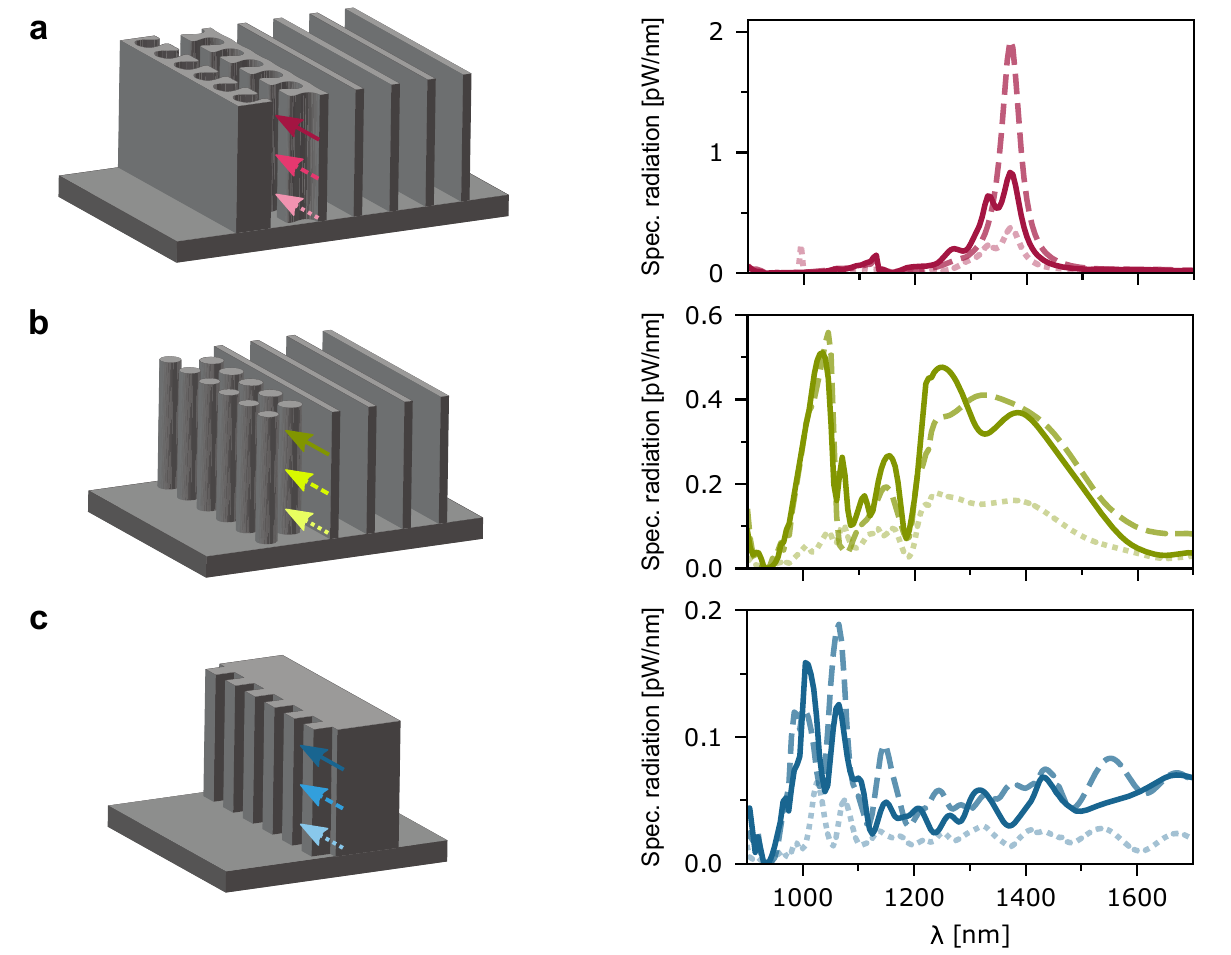}
  \caption{\textbf{Dependence on beam height for all three structure designs}. The radiation spectra were calculated from 3D-FD simulations for different heights of the electron beam, where the beam positions were measured from the bottom of the 1.5-\SI{}{\micro m}-high nanostructures: \SI{0.25}{\micro m} (dotted), \SI{0.75}{\micro m} (dashed) and \SI{1.25}{\micro m} (solid). \textbf{a}, Inverse design. \textbf{b}, Dual pillar with DBR. \textbf{c}, Rectangular grating. All three designs exhibit lower emission when the beam is \SI{0.25}{\micro m} close to the substrate. The dual pillar structure and rectangular grating show little difference between middle and high positions, whereas the inverse design performs substantially better with the beam focused at central height (\SI{0.75}{\micro m}). This demonstrates a resonating behavior of the inverse design structure, strongest at the center.} \label{figED2}
\end{figure*}

\begin{figure*}
  \centering
  \includegraphics{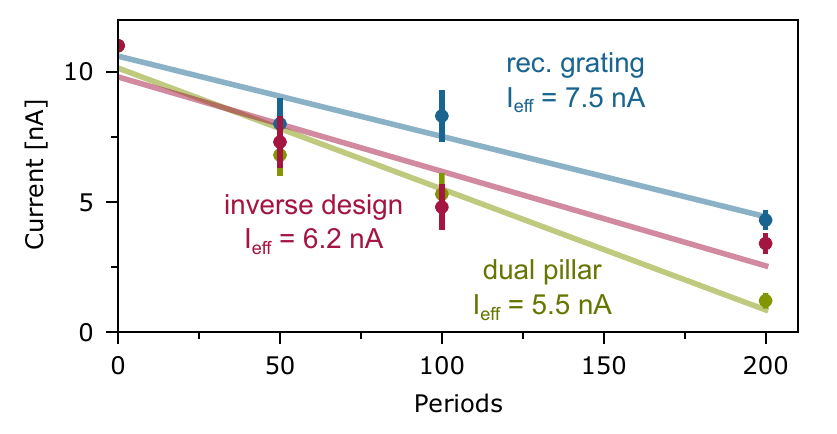}
  \caption{\textbf{Determination of effective current.} The electron beam current deteriorates as part of the beam collides with the silicon nanostructures. The current was measured with a Faraday cup before and after passage through different structure versions with 50, 100 and 200 periods. The beam loss is smallest for the single-sided, rectangular grating (blue) and largest for the dual pillars (green), which have a smaller channel width than the inverse design (red). From a linear fit, we determined the effective current $I_\mathrm{eff}$ as the mean current within the structure or correspondingly the current after passage through half the structure.} \label{figED3}
\end{figure*}

\begin{figure*}
  \centering
  \includegraphics{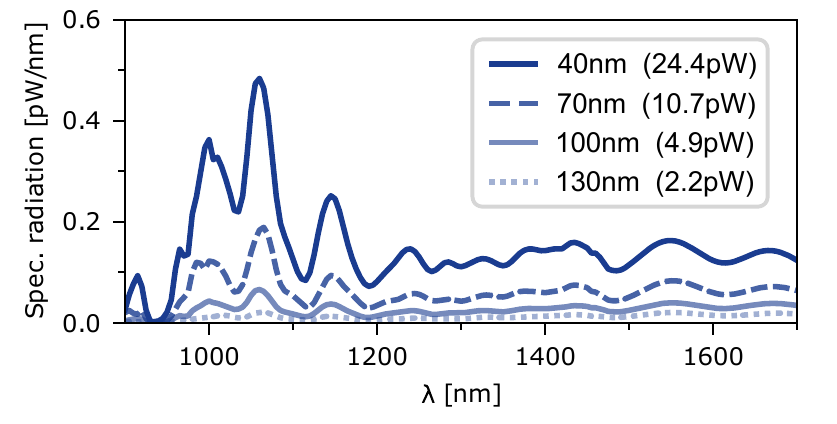}
  \caption{\textbf{Dependence on beam-grating distance}. Simulated radiation spectra of the rectangular grating are shown for various beam-grating distances ($40\,\mathrm{nm}$, $70\,\mathrm{nm}$, $100\,\mathrm{nm}$, $130\,\mathrm{nm}$). The 3D-FD simulation assumed a current of $11\,\mathrm{nA}$ and a beam width of $\sigma = 20\,\mathrm{nm}$. The power values in brackets, obtained by integration over the Gaussian collection range, demonstrate high dependence of the efficiency on the beam distance. We fitted an exponential function to these power values to estimate the beam-grating distance used in the experiment. With an effective current of $7.5\,\mathrm{nA}$, we found that the actual distance in the experiment was $58\,\mathrm{nm}$.} \label{figED4}
\end{figure*}

\begin{figure*}
  \centering
  \includegraphics{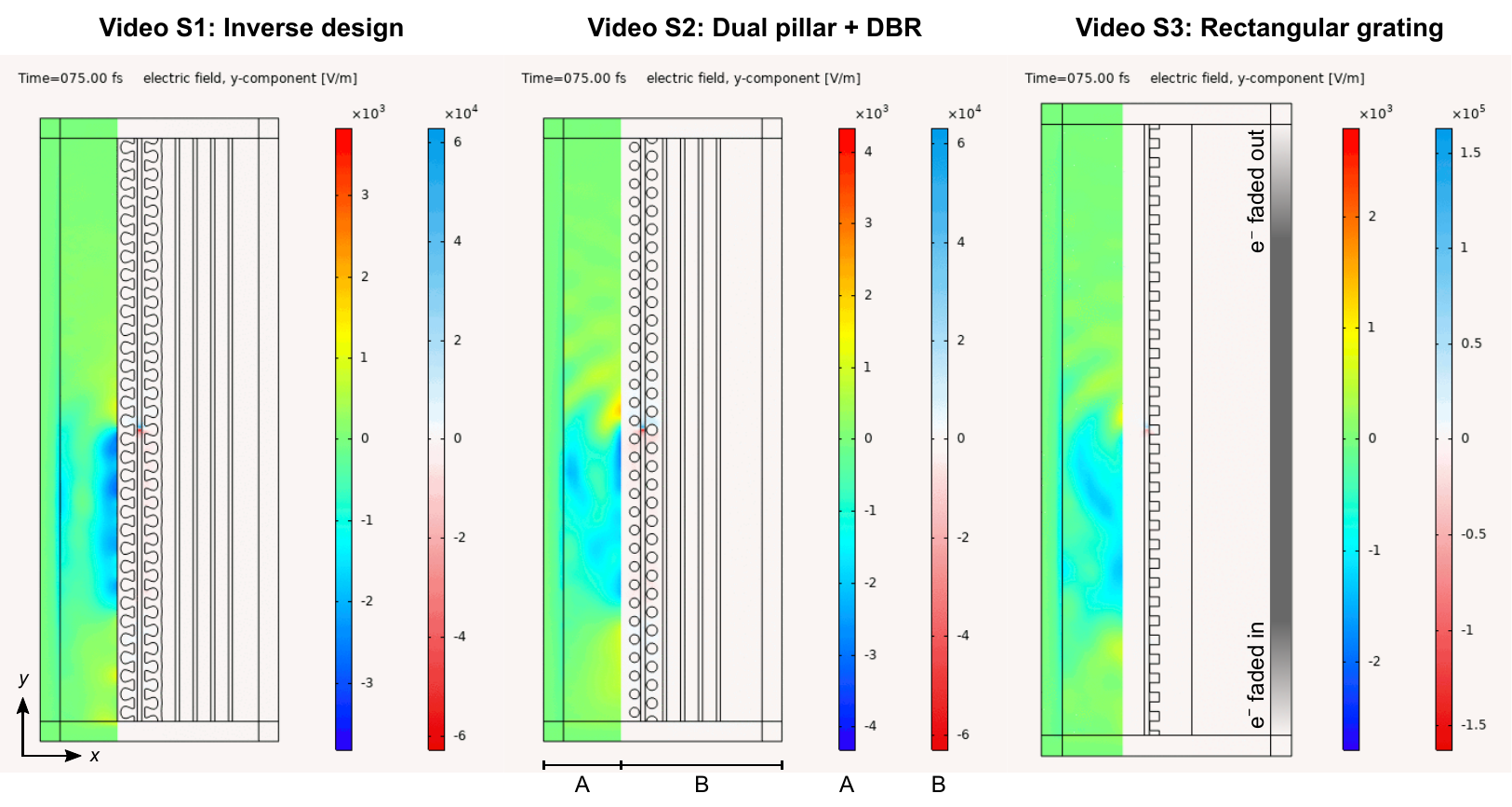}
  \caption{\textbf{2D time-domain simulations}. The three videos show the emission from a single electron traversing the different structures. Each structure is 32 periods long. To avoid numerical errors, the electron is slowly faded in after the 6\textsuperscript{th} period ($t \approx 20\,\mathrm{fs}$) and faded out after the 26\textsuperscript{th} period. The electric field $E_y$ in direction of the electron propagation is visualized with different colormaps for the near field (red-blue) and the far field (rainbow).}
\end{figure*}

\end{document}